\documentclass[12pt,tightenlines,eqsecnum,floats,aps,amsmath,amssymb,nofootinbib,prd,superscriptaddress]{revtex4}

\usepackage{setspace}
\usepackage{amsmath,amssymb,amsfonts,amsthm,mathrsfs}
\usepackage{graphicx}
\usepackage{enumerate}

\def\be{\begin{equation}}
\def\ee{\end{equation}}
\def\ba{\begin{eqnarray}}
\def\ea{\end{eqnarray}}
\def\bi{\begin{itemize}}
\def\ei{\end{itemize}}

\def\tr{\text{Tr}}
\def\O{\Omega}
\def\Oc{\Omega^c}
\def\F{\cal F}
\def\R{\cal R}
\def\S{\cal S}
\def\G{\sqrt{8\pi G}}
\def\bra{\langle}
\def\ket{\rangle}

\begin{document}

\title{A Note on Entanglement Entropy, Coherent States and Gravity}
\author{Madhavan Varadarajan}\email{madhavan@rri.res.in}
 \affiliation{Raman Research Institute, Bangalore 560080, India}

\begin{abstract}

The entanglement entropy of a free quantum field in a coherent state is 
independent of its stress energy content. 
We use this result to highlight the fact that 
while the Einstein equations for first order variations about a locally maximally symmetric
vacuum state of geometry and quantum fields seem to follow from Jacobson's principle of 
maximal vacuum entanglement entropy, their possible derivation from this principle for 
the physically relevant case of finite but small variations 
remains an open issue.
%
We also apply this result to 
the context of Bianchi's identification, independent of unknown Planck scale physics,
 of  the first order variation of 
Bekenstein- Hawking area  with that of vacuum 
entanglement entropy. We argue that under certain technical assumptions this identification  seems not to be extendible
 to the context of 
finite but small variations to coherent states.

Our particular method of estimation of entanglement entropy variation reveals the existence of certain contributions
over and above those of References \cite{ted,bianchi}. We discuss the sense in which these contributions may be subleading
to those in  References \cite{ted,bianchi}.


\end{abstract}
\maketitle

\section{Introduction} \label{sec1}

In his path breaking  work of 1995, Jacobson \cite{tedold} 
derived the classical Einstein equations as conditions of thermal
equilibrium in local Rindler Wedges at every point in spacetime. 
Recently, Jacobson \cite{ted} provided a deeper implementation of his ideas by 
moving his analysis from the 
setting of local Rindler wedges  to causal domains  of  small spherical regions about every spacetime
point. Instead of {\em classical} matter flux through  local Rindler horizons, he considered {\em quantum} matter fields
restricted to these causal domains and thereby obtained the  {\em semiclassical} Einstein equations
as equilibrium conditions on the entanglement entropy of the quantum
 matter in these causal domains. More in detail, Jacobson proposed that this entanglement entropy 
is maximal when the local geometry in the causal domain is maximally symmetric and the quantum fields are in the 
associated maximally symmetric vacuum. This implies that first order variations about this maximal entropy 
configuration must vanish where, by first order variation we mean the derivative with respect to a parameter
labelling the state and geometry, 
evaluated at the maximally symmetric configuration.
Jacobson showed that this implication of maximal vacuum entropy
was equivalent to the satisfaction of 
the Einstein equations for  such first order variations. 
A key feature of his analysis is the emergence of 
the stress energy 
of the matter field 
through quantum entanglement via the so called First Law of Entanglement Entropy\cite{myersetal}. 
This Law, under certain conditions \cite{ted,myersetal},
relates first order variations
of stress energy expectation value to those of entanglement entropy.

Whereas first order variations  involve limiting behaviour under infinitesmal changes,
for applicability to  physical situations one would hope that the analysis of Reference \cite{ted} could be 
extended from the case of 
infinitesmal variations to finite but small variations.
In this note we recall that there exist, at least for free field theories,  quantum states which are finite but
small variations of the vacuum and 
whose   entanglement entropy is {\em independent} of their stress energy content. This fact seems to be in tension 
with  the non-trivial dependence of stress energy on entropy suggested by the First Law of Entanglement Entropy, and 
hence with Jacobson's results. 
As we shall see, the  resolution of this apparent tension  is that for 
coherent states which are  finite but small  variations about the vacuum, the stress energy expectation value 
 vanishes to first order in the smallness parameter.
Thus, to first order in the smallness parameter, the variation of the entropy and that of the stress energy both vanish.
Since the First Law of Entanglement Entropy as well as Jacobson's derivation are  first 
order analyses, no contradiction arises. 
Nevertheless, as the work of Ford and Kuo \cite{fordkuo} indicates, 
quantum matter in a coherent state provides a setting in which we {\em do} expect the 
semiclassical Einstein equations to hold irrespective of the smallness of the stress energy content of the coherent state.
Hence a generalization of the considerations of Reference \cite{ted} to coherent states which are finite but small
variations of the vacuum constitutes an (extremely
 interesting) open issue. 

Our general remarks pertain to any setting in which the First Law of Entanglement Entropy is used to
derive  gravitational consequences of 
entanglement entropy variation. Another such setting is 
  that of Bianchi's intriguing derivation of the first order variation of Bekenstein- Hawking entropy 
from the first order variation of entanglement entropy of quantum states across Causal Horizons \cite{bianchi}
(see also Bianchi and Satz \cite{bianchisatz}) without recourse to UV physics.
Once again, for  applicability to physical
situations we would like to enquire as to whether the considerations of Reference \cite{bianchi} also hold
for  finite but small variations. In this regard we shall argue that 
this derivation (modulo some technical assumptions) does not seem to  be 
extendible  to the context of coherent states which constitute finite but small variations of the 
vacuum.


The main purpose of this work is to detail the observations  of the previous paragraphs.
For concreteness we shall restrict attention to matter constituted by a single free scalar field.
The layout of this note is then as follows.
In section 2 we recall that the entanglement entropy of any coherent 
state of a free quantum field in a causal domain  is the same as the entanglement entropy of the 
free field vacuum state in the causal domain. 
Next,  we show that in a 
 coherent state 
which is a  finite, small variation
about the vacuum, the stress energy expectation value is non-trivial only at second order in the smallness parameter.

In section 3 we 
compute the first order variation of entanglement entropy and stress energy under
variations of geometry and state around their maximally symmetric vacuum configurations. 
Our particular method of computation yields certain contributions to these variations over and above 
those which appear in Reference \cite{ted}. We discuss the sense in which these contributions maybe 
subleading to those in \cite{ted}. From the considerations of \cite{ted}, it then follows that 
their neglect reproduces the derivation of the Einstein 
equations for such (infinitesmal) variations.

In Section 4  we estimate  the change in entanglement entropy $S$ and stress energy $T_{ab}$ 
under {\em finite} but {\em small} variations 
of geometry and state. More in detail, we consider a 1 parameter family of geometries and states and estimate 
$S,T_{ab}$ at finite, small values of the parameter relative to their values at the origin in the context of a Taylor
approximation.
We use the results
of Sections 2 and  4 in sections 5 and 6. Hence our conclusions in those sections are contingent
on our assumed existence of the Taylor approximation of section 4.

In  section 5 we confront Jacobson's principle of maximal vacuum entanglement entropy \cite{ted} with 
coherent states which are finite but small variations of the vacuum.
We argue that while the existence of a 
{\em derivation} of the Einstein equations  from Jacobson's principle of maximal vacuum entanglement entropy
for variations to such coherent states remains an open issue,
the Einstein equations may be {\em consistent} with this principle.
\footnote{This consistency was anticipated by Jacobson in \cite{ted}}

In section 6, we turn our attention to the considerations of Reference \cite{bianchi} and argue that 
(under certain technical assumptions)
Bianchi's  ideas relating first order variations of Bekenstein Hawking area with those of entanglement entropy
using the quantum Einstein equations do not seem to be
extendible  to the case of coherent states
 which are finite but small  variations of the vacuum.



\section{Entanglement Entropy and Stress energy of Coherent States.}

The result  that the entanglement entropy of any coherent state is the same as the entanglement entropy
of the vacuum state has been derived in various contexts 
\cite{fiola,benedict,das}. 
While Reference \cite{fiola}
focuses on 1+1 dimensions and uses the conformal flatness of two dimensional geometries, 
the considerations of References \cite{benedict,das} are restricted to flat spacetimes.
As in References \cite{fiola,benedict,das}, we shall restrict attention in this paper to free scalar field
matter.

The underlying reason for the equality of coherent state and vacuum entanglement entropies
as succintly stated in Reference \cite{benedict} is that any coherent state can be written as a 
particular unitary transformation of the vacuum which factorizes as the product of  a pair unitary transformations
one depending only on the degrees of freedom being traced over and the other, on the remaining degrees of freedom.
The equality of entanglement entropies then follows from the invariance of the tracing operation under such unitary maps.
More in detail, 
consider a free (real) scalar field $\Phi$ on a
4 dimensional flat spacetime with inertial coordinates $({\vec x},t)$. 
Consider a coherent state $|H\ket$ which is modelled on 
a classical scalar field solution $H({\vec x},t)$. It is straightforward to verify that
\be
|H\ket = {\hat U}(e,f)|0\ket ,
\label{mcoh}
\ee
where 
\be
{\hat U}(e,f):= e^{i\int_{R^3} d^3x f({\vec x}){\hat \phi} ({\vec x}) - e({\vec x}) {\hat \pi} ({\vec x})}.
\ee
Here, the integral is over some constant $t$ slice, ${\hat \phi} ({\vec x}), {\hat \pi} ({\vec x})$ are the scalar field and conjugate momentum operators
on this  slice and $e({\vec x}), f({\vec x})$ are the scalar field and conugate momentum induced
on the slice by the spacetime solution $H({\vec x},t)$.

Let $\O$ be a 3 dimensional region
on this slice. The entanglement entropy of a state is the Von Neumann entropy of the 
density matrix obtained by tracing 
over the scalar field degrees of freedom in the complement of  $\Omega$. Let us label quantities defined  within $\Omega$ 
by the subscript $\O$ and those defined in $R^3- \Omega$ by $\Oc$. For example, $\tr_{\Oc}$ refers to
a trace over  degrees of freedom in $\Oc$.
The density matrix for $|H\ket$ is then ${\hat \rho}(H) = \tr_{\Oc}(|H\ket \bra H|)$  
and that for the vacuum is  ${\hat\rho} (0)= \tr_{\Oc}(|0\ket \bra 0|)$. 
Denote the corresponing  entanglement entropies by 
$S(H)= -\tr_{\O}({\hat \rho}(H)\ln {\hat \rho}(H))$ , 
$S(0)= -\tr_{\O}({\hat \rho}(0)\ln {\hat \rho}(0))$.

We have the factorization:
\be
{\hat U}(e,f)= 
{\hat U}_{\Oc}(e,f){\hat U}_{\O}(e,f)
\label{mu1u2}
\ee
with 
\be 
{\hat U}_{\O}(e,f)  =e^{i\int_{\O} f (x)\phi (x) -e(x) \pi (x)},\;
{\hat U}_{\Oc}(e,f)  =e^{i\int_{\Oc} f (x)\phi (x) -e(x) \pi (x)}.
\label{exp}
\ee
It follows that
\ba
{\hat \rho}(H) &=& \tr_{\Oc} ({\hat U}_{\Oc}(e,f){\hat U}_{\O}(e,f)|0\ket\bra 0|
             {\hat U}_{\O}(e,f)^{\dagger}{\hat U}_{\Oc}(e,f)^{\dagger})\\
&=& \tr_{\Oc}({\hat U}_{\O}(e,f)|0\ket \bra 0|
             {\hat U}_{\O}(e,f)^{\dagger}) = 
{\hat U}_{\O}(e,f)\tr_{{\Oc}}(|0\ket \bra 0|){\hat U}_{\O}(e,f)^{\dagger},
\label{murhou}
\ea
where we have used the invariance of the Trace under unitary tranformations, and the fact that $U_{\O}(e,f)$ acts
as the identity operator on the degrees of freedom in $\Oc$ in the last  line. The equality of $S(H)$ and $S(0)$ then follows, once again,
from the invariance of the Trace under unitary transformations.

While equations (\ref{mu1u2})- (\ref{murhou}) exhibit the simple underlying reason for the equality of $S(H)$ and $S(0)$,
the equations themselves are UV divergent and need to be regulated because of the sharp boundary between $\O$ and $\Oc$.
The sharp boundary renders the calculations UV divergent in two ways.
The first is the well known 
UV divergence of $S(0)$ itself. The second is that for  $(e,f)$ which are non-vanishing on this boundary,
the exponents in equation (\ref{exp}) are not well defined self adjoint operators.
For concreteness let us use a lattice regulator as in Reference \cite{das} to regulate the first divergence.
The second one then automatically disappears because the integrals of quantum fields over regions 
in the exponents of (\ref{exp}) become sums of quantum mechanical degrees of freedom over lattice sites and the 
exponents  are thereby rendered mainfestly well defined. Since both divergences spring from the sharpness of the 
boundary, we expect that any UV regulated calculation which effectively removes the sharpness of this boundary
makes  equations (\ref{mu1u2})- (\ref{murhou}) well defined and establishes the equality of $S(H)$ and $S(0)$.

Next we turn to the (flat spacetime) normal ordered stress energy tensor.
It is easy to check that the normal ordered two point function in the coherent state $|H\ket$ evaluates to
\be 
\bra H| :{\hat \Phi}({\vec x}, t){\hat \Phi}({\vec x}^{\prime}, t^{\prime}):|H\ket=
H({\vec x},t)H({\vec x}^{\prime},t^{\prime} )
\label{mhad}
\ee
and the stress energy tensor expectation value is then exactly that of the classical solution $H({\vec x},t)$.
We are interested in coherent states which are first order departures from the vacuum. 
Since the free field equations are linear, if $H$ is a solution with initial data $(e,f)$ as above, then 
$\delta H= \delta \times H$ is a solution with initial data  $(\delta e, \delta f)$.
Accordingly for 
some small positive parameter $\delta$ we have
\be
 |\delta H\ket = e^{i\delta \int_{R^3}d^3x f({\vec x}){\hat \phi} ({\vec x}) - e({\vec x}) {\hat \pi} ({\vec x})}|0\ket .
\label{mdef}
\ee
It is straightforward to expand the exponential out and verify that this state is a first order (in $\delta$) variation
of the vacuum state as desired. It also follows from equation (\ref{mhad}) that the normal ordered 
stress energy expectation  value is of second order in $\delta$.

Our considerations so far have been in the context of flat spacetimes. In the appendix we define coherent states 
on curved spacetimes through an appropriate generalization of equation (\ref{mcoh}) and derive the curved spacetime
generalizations of equations (\ref{mu1u2})- (\ref{mdef}). While a detailed regulation of  the UV divergences 
in the curved spacetime analogs of equations (\ref{mu1u2})- (\ref{murhou}), is out of the scope of this work,
the general discussion above  of the softening of the sharp boundary via UV regulation indicates that  
the equality of vacuum and coherent state entanglement must continue to hold in the curved spacetime context.

\section{First order variation of entanglement entropy and stress energy}
The entanglement entropy $S$ for degrees of freedom restricted to the spatial region $B$
and the stress energy expectation value $T_{ab}$ on $B$
 depend on the spacetime geometry $g$ and the quantum state of the field $\psi$ in the vicinity of $B$. We 
indicate this dependence of $S,T_{ab}$ through the notation $S(g,\psi ), T_{ab}(g,\psi )$.
Let $\lambda$ be a real parameter taking values in some small interval about the origin.
Let $B(\lambda )$ be the spatial region of interest. Let $g(\lambda )$ be the spacetime metric in
the vicinity of $B(\lambda )$ and let $\psi (\lambda )$ be the free scalar field state on $B(\lambda )$.
In Reference \cite{ted}, $B(\lambda )$ denotes a small fixed volume geodesic ball (with respect to the
metric $g(\lambda )$) around a spacetime point $p$. We use the diffeomorphism freedom to identify
$B(\lambda )$ with $B(0):=B$. Finally, let $(\psi (0), g(0))$ be the maximally symmetric vacuum configuration.


The first order variation of any functional  $A(g, \psi )$ is defined to be
$\left(\frac{d}{d\lambda} A(g(\lambda), \psi (\lambda ) )\right)|_{\lambda=0}$
The computation of the derivative assumes that there is some way of comparing the pair 
$(g(\lambda), \psi (\lambda )$ with the pair $(g(0), \psi (0))$ in an arbitrarily small neighbourhood of the origin.
Since we are not aware of an explicit way of making such a pairwise comparision, we shall assume that 
we can {\em independently} vary $g$ and $\psi$ in the computation of this derivative.
Thus we compute the derivative as a sum of two {\em partial} derivatives, one obtained by holding the geometry $g$ fixed 
at $g(0)$ and varying the state $\psi$ and the other by holding the state $\psi$ fixed at $\psi (0)$ and varying 
the geometry $g$. This in turn requires that we are able to compare the state $\psi (\tau )$ with $\psi (0)$
{\em independent} of the geometry $g (\lambda )$ and vice versa when $\lambda , \tau$ vary over some arbitrarily 
small neighbourhood of the origin. In particular, this implies that  we are able to identify the {\em same} state on
different  geometries in an arbitrarily
 small neighbourhood of parameter space.
We shall assume that it is possible for this identification to be made in such a way that the fixed state
is Hadamard on these different geometries.

More precisely, we consider a 2 parameter family of states and geometries $(g(\lambda ),\psi (\tau ))$
with $\psi (\tau )$ assumed to be 
Hadamard with respect to the spacetime geometry  $g(\lambda )$ in the vicinity of $B$ and 
in a small neighbourhood of parameter space around $\lambda=0=\tau$.
It then follows that the first order variations of $S, T_{ab}$ evaluate to
\be
\left(\frac{dS(g(\lambda), \psi (\lambda ))}{d\lambda}\right)_{\lambda=0}=
  \big(\frac{\partial S(g(\lambda), \psi (0 ))}{\partial \lambda\;\;\;\;}\big)_{\lambda=0}
+  \big( \frac{\partial S(g(0), \psi (\tau ))}{\partial \tau \;\;\;\;} \big)_{\tau=0} 
\label{s1st}
\ee
\be
\left(\frac{dT_{ab}(g(\lambda), \psi (\lambda ))}{d\lambda}\right)_{\lambda=0}=
 \big(\frac{\partial T_{ab}(g(\lambda), \psi (0))}{\partial \lambda\;\;\;\;}\big)_{\lambda=0}
+ \big( \frac{\partial T_{ab}(g(0), \psi (\tau ))}{\partial \tau \;\;\;\;} \big)_{\tau=0} 
\label{t1st}
\ee

Jacobson's proposal of maximal vacuum entanglement implies that the right hand side of equation(\ref{s1st}) vanishes.
The second term  in equation (\ref{s1st}) concerns the contribution 
obtained by varying the state but keeping the geometry fixed.
Jacobson refers to this term  as the IR contribution because it is independent of the UV physics which regulates the
divergence in the entropy arising from short distance correlations close to the entangling surface.
The assumed Hadamard behaviour of all states considered implies that
the fixed geometry contribution of the second term is expected to be finite and dependent
only on the IR physics. This is indeed true and the First Law of Entanglement entropy reexpresses this term
as $1/\hbar$ times (an integral of) the second term of equation (\ref{t1st}).
\footnote{This is true for conformal matter, for the non-conformal case we assume Jacobson's conjecture \cite{ted} holds.
See also Reference \cite{myers} in this regard.}

The first term in equation (\ref{s1st}) corresponds to the contribution 
from the fixed state while varying the geometry. This is called
the  UV term in \cite{ted}. Since the area of the entangling surface changes when the metric is varied,
and since the UV divergence of the entropy is expected, to leading order in the UV cutoff, to scale as 
the area one assumes that this term has a UV contribution corresponding to the variation of the area.
However, in principle,
there may also be additional IR contributions to the entropy
which depend on the precise identification of the unvaried (vacuum) state as a state on the varied geometry.

If these IR contributions can be neglected, the argumentation of \cite{ted} holds whereby
 the UV contribution is related to
the $1/\hbar$ times `$G_{ab}$' term and the remaining IR contribution, as mentioned above, 
to $1/\hbar$ times the second term in the `$T_{ab}$' equation (\ref{t1st}).
The Einstein equations for first order variations 
would then ensue if we could neglect the first term in equation (\ref{t1st})
compared to the second term in that equation.
Once again whether we can do this depends on the manner in which 
the unperturbed state is identified as a state on the perturbed spacetime. 
To summarise: If we assume that the geometry and state can be varied independently, 
we are able to obtain the Einstein equations for first order variations 
from equations (\ref{s1st}) and (\ref{t1st}) in conjunction with the principle of maximal vacuum
entanglement only if 
the identification of the unvaried state on the varied geometry is such that
we are able to neglect the first term of (\ref{t1st}) as well as the 
IR contribution to the first term of (\ref{s1st}).

Recall that the need to identify state spaces for the varied and unvaried geometries stems from our (current) inability
to make a pairwise comparion between $(g (\lambda ), \psi (\lambda ))$ and $(g(0), \psi (0))$. If we could make 
such a pairwise comparision, the need to identify $\psi (0)$ as a state on the geometry $g(\lambda )$ would not be
necessary. This suggests that our final conclusions should be independent of the way in which we identify
state spaces on the varied and unvaried spacetime geometries.
We also note that there is no canonical way to make this identification for generic variations of the geometry
due to the generic lack of any (conformal) symmetry of the varied geometry. This suggests that we 
try to make this identification in such a way that the first term of (\ref{t1st}) and
the IR contribution to the first term of (\ref{s1st}) 
can indeed be neglected.

In order to see one particular way in which 
these sort of `identification dependent'  contributions could be negligible, we present an argument
suitably adapted here from Hawking\cite{hawking} which suggests that for {\em any} physically reasonable 
identification  the first term in equation (\ref{t1st}) can be neglected compared to the second.
The argument is as follows.
%
The metric $g(\delta )$ at parameter value $\lambda = \delta$ is a small variation of $g(0)$ for $\delta <<1$
and we write it as $g(\delta )= g (0)+ \delta g$.
For  modes which are high frequency with respect to the curvature scale, we do not expect a significant change in 
the annihilation- creation operators on  the spacetime geometry $g(0)+\delta g$ relative to the annihilation- creation 
operators on $g(0)$.
This is consistent with the assumed Hadamard behavior of the states.
For modes which have a frequency comparable or less than the
curvature scale, we expect  that the annihilation- creation operators 
for the spacetime $g(0)+\delta g$ are modified by order $\delta$
relative to the annihilation- creation operators for $g(0)$. This results in a modification of order $\delta$ in the
number operator for these modes. Approximating the mode density by the flat spacetime mode density 
$\omega^2 d\omega$  of the number of modes per unit volume at frequency $\omega$, we obtain a 
particle number density of the order of $\delta L^{-3}$ where the curvature scale of $g$ is  $L^{-2}$. 
This in turn leads to an   uncertainty of the energy per unit volume of order $\hbar \delta L^{-4}$,
which is first order in $\hbar$. This argument (in the limit that  $\delta \rightarrow 0$) suggests that 
the first term in equation (\ref{t1st}) is  of order $\hbar$. Since the stress energy in the Einstein equations
is expected to be independent of $\hbar$, the first  term of (\ref{t1st}) is  subleading in $\hbar$ and can 
be neglected.

It would be good to have  a similar argument for the neglect of any IR contributions to the first term in equation (\ref{s1st}).
The argument above indicates that   the uncertainty in the number of particles per unit volume  is $\delta L^{-4}$
(which is  of order $\hbar^0$)
but  we are not sure as to how
this could be used to conclude that  the additional IR corrections to the entropy  arising from this first term 
are also of  order $\hbar^0$.

\section{Finite variations of entanglement entropy and stress energy}

In this section we estimate the change in $S, T_{ab}$ under joint  variations of geometry and state which are
finite and small. Thus we are interested in the estimation of 
$S(g (\delta  ), \psi (\delta ))- S(g(0), \psi (0) )$ and 
$T_{ab}(g (\delta  ), \psi (\delta ))- T_{ab}(g(0), \psi (0) )$ for $\delta <<1$.
Similar to the previous section, we assume that the metric and state can be independently varied and  that 
for $\tau, \lambda$ in a small neighbourhood of the origin 
an identification, of the state $\psi (\tau )$ on the spacetime geometry $ g(\lambda )$ in the vicinity of the spatial
region of interest $B$,
has been made in such a way as to  preserve its Hadamard property. 
Further, we shall assume that $S,T_{ab}$ admit a Taylor approximation to second order in the smallness parameter $\delta$.

The expansion  of the entanglement entropy to second order in $\delta$ is then:
\ba
S(g (\delta  ), \psi (\delta ))- S(g(0), \psi (0) )& =&
\Delta S^{(1)} +\Delta S^{(2)} \label{s1s2}\\
\Delta S^{(1)}:= \delta  \big(\frac{\partial S}{\partial \lambda\;\;\;\;}\big)_{\lambda=0}
&+& \delta \big( \frac{\partial S}{\partial \tau \;\;\;\;} \big)_{\tau=0} 
\label{s1} 
\\ 
\Delta S^{(2)}:= 
\frac{ \delta^2}{2}\big( \frac{\partial^2 S}{\partial \lambda^2}\big)_{\lambda=\tau=0}
&+&\frac{ \delta^2}{2}\big( \frac{\partial^2 S}{\partial \tau^2} \big)_{\lambda=\tau=0}
+ \delta^2\big( \frac{\partial^2 S}{\partial \lambda \partial \tau} \big)_{\lambda=\tau=0}, 
\label{s2}
\ea
and that for the stress energy is:
\ba
T_{ab}(g (\delta  ), \psi (\delta ))- T_{ab}(g(0), \psi (0) )& =&  
\Delta T_{ab}^{(1)} +\Delta T_{ab}^{(2)}
\label{t1t2}
\\
\Delta T_{ab}^{(1)}:=  
\delta\big(\frac{\partial T_{ab}}{\partial \lambda\;\;\;\;}\big)_{\lambda=0}
&+ &\delta\big( \frac{\partial T_{ab}}{\partial \tau \;\;\;\;} \big)_{\tau=0} 
\label{t1}
\\ 
\Delta T_{ab}^{(2)}:=
\frac{\delta^2}{2}\big( \frac{\partial^2 T_{ab}}{\partial \lambda^2}\big)_{\lambda=\tau=0}
&+&\frac{\delta^2}{2}\big( \frac{\partial^2 T_{ab}}{\partial \tau^2} \big)_{\lambda=\tau=0}
+\delta^2 \big( \frac{\partial^2 T_{ab}}{\partial \lambda \partial \tau} \big)_{\lambda=\tau=0}
\label{t2}
\ea

Equations  (\ref{s1}),  (\ref{t1})
constitute the  first order contribution in the smallness parameter $\delta$ to the variation of $S, T_{ab}$ so that the
right hand side of these equations is identical to that of equations (\ref{s1st}) and (\ref{t1st}) multiplied by 
$\delta$.

Equations (\ref{s2}),  (\ref{t2}) list the second order contributions.
We note that terms which have no partial derivative(s) with respect to $\tau$ correspond to contributions from 
varying the geometry while keeping the state fixed. Similarly, terms which have no 
partial derivative(s) with respect to $\lambda$ correspond to contributions from 
varying the state while keeping the geometry fixed.

Motivated by the discussion  at the end of Section 3, we shall assume, for the rest
of the paper that it is possible to identify state spaces on different geometries in such a
way that the following holds:\\

\noindent {\bf A1}:{\em All `IR' contributions to $S,T_{ab}$ in equations (\ref{s1})- (\ref{t2}) arising from 
 varying the geometry while keeping the state fixed  can be neglected.}

\section{Comments related to coherent states in the  setting of Reference \cite{ted}}

As mentioned in section 1, the emergence of the Einstein equations for first order variations from the proposal of maximal
vacuum entanglement suggests that we enquire if this principle has anything to say about 
the Einstein equations for the physically interesting case of 
{\em finite} but small variations to coherent states. In this section we pursue this question using the estimates
of the variation of $S,T_{ab}$ displayed in the previous section in equations (\ref{s1})- (\ref{t2}) for variations to 
coherent states described by equation (\ref{def}).

First consider the estimate to first order in the smallness parameter $\delta$ given by equation (\ref{s1}).
Specializing to the 
coherent state case, the fact that on a fixed geometry the entanglement entropy of a coherent state is 
the same as that of the vacuum implies that the second term on the right hand side of equation (\ref{s1})
vanishes.
As in section 3, the First Law of Entanglement entropy relates this term to the 
second term of equation (\ref{t1}). From section 2 and the Appendix we see that this term also 
vanishes by virtue of the fact that the stress energy expectation value at fixed geometry 
is quadratic in the parameter $\delta$.
Thus there is no contradiction between the First Law of Entanglement entropy and the fact that the entanglement entropy
of a coherent state is independent of its stress energy.

Note also that Jacobson's proposal of maximal vacuum entanglement entropy implies, as discussed in section 3, 
that the right hand side of (\ref{s1}) vanishes under the assumption {\bf A1} of section 4.
This, in turn, implies the vanishing of the UV contribution
  to the first term of  (\ref{s1}). This implies that the area of the small geodesic ball under consideration 
\footnote{At this stage it is appropriate to note that our considerations in this section  
are based on a single small geodesic
ball centred on some fixed spacetime point and on the unperturbed maximally symmetric geometry of this ball,
its associated vacuum state and coherent state excitations of this vacuum.
In contrast, the principle of maximal vacuum entanglement applies to {\em any} spacetime point, the vacuum
in question then being the local vacuum associated with the local patch of maximally symmetric spacetime
about this point. In Reference \cite{ted}, while the unvaried state is this {\em local} vacuum, the
varied state is {\em global} in the sense that it is defined on the entire spacetime not just a local patch.
If we choose this global  varied state such that its restriction to a fixed ball about a fixed spacetime point is
that of  a coherent state excitation of the
{\em local} vacuum in that ball, the extent to which we can continue to think of this global state as a 
coherent state excitation of the 
local vacuum as the spacetime point varies over the entire spacetime , is not clear.}
does not change to first order in the smallness parameter $\delta$.
Finally, 
under the assumption {\bf A1}, 
the first term on the right hand side of (\ref{t1}) also vanishes. Thus we conclude that the change in  entanglement 
entropy  and  stress energy expectation value under finite but small variations to coherent states vanish
to first order in the smallness parameter.
This is consistent with the fact that under such variations (and under assumption {\bf A1}),
the Einstein equations relate the  non-trivial area change of the small geodesic ball to the stress energy tensor
only at second order in the smallness parameter. 

Let us now move on to the contributions to $S,T_{ab}$ at second order in the smallness parameter
as catalogued in (\ref{s2}) and (\ref{t2}). While we would like to use these equations to 
derive the Einstein equations for  finite, small variations to coherent states 
from the principle of maximal vacuum entanglement, we are unable to do so. Instead 
we shall use these equations to examine 
if  
the Einstein equations
are {\em consistent} with the principle of maximal vacuum entanglement entropy for such variations.

The first term in equation (\ref{t2})  
as well as the IR contribution to the 
first term in equation (\ref{s2})  
can be neglected by virtue of our assumption {\bf A1}.
The second term in these equations contains no $\lambda$ derivative and corresponds to the 
contribution when the state is  varied and the geometry is held fixed.
In the coherent state case these contributions can be evaluated: in  (\ref{s2}) this term  vanishes by virtue of 
our considerations in section 2 and the Appendix,  and  
in (\ref{t2}) this term yields the stress energy of the coherent state 
on the unperturbed geometry. From equation (\ref{had}), this term in (\ref{t2}) is exactly the classical stress energy
of the classical free scalar field solution which labels the coherent state. We assume that the classical stress
energy satisfies the weak energy condition so that this term is positive.

Finally, we also have the third `mixed' terms in equations (\ref{s2}) and (\ref{t2})
and we do not know at present
how to estimate them. If these terms are negligible then the only remaining contribution to (\ref{s2}) 
is the UV term and the only remaining contribution to (\ref{t2}) is its second term, which as argued above, is
positive. Further, since (under assumption {\bf A1}) the contributions (\ref{s1}), (\ref{t1}) at first order 
in the smallness parameter vanish, the leading order contribution to the entanglement entropy (\ref{s1s2})
is the UV term at second order in the smallness parameter and the leading order contribution to the 
stress energy (\ref{t1t2})  is the second term of (\ref{t2}) which is positive and also of  second order.

Let us now assume that the Einstein equations hold. From Reference \cite{ted} it follows that the 
positive stress energy leads to an 
area deficit. Since the UV contribution in (\ref{s2}) is proportional to this area 
variation, this UV contribution  is {\em negative}.
Thus if the terms we have neglected can indeed be neglected and if the Taylor approximation is valid, 
we find that the variation of the total entanglement entropy 
is bounded by a  {\em negative} quantity times $\delta^2$
so that  the maximal entanglement entropy principle is {\em consistent} with the Einstein 
equations in the coherent state case as anticipated by Jacobson \cite{ted}.

Before we end this section we would like to emphasise that coherent states can only be defined in the 
context of {\em free} field theory; our considerations would therefore {\em not} apply to (strongly) interacting 
quantum fields. 

\section{Comments related to the setting of Reference \cite{bianchi}}

The proposal that 
Black Hole entropy be identified with entanglement entropy of quantum fields across the horizon 
was first made by Sorkin and his collaborators \cite{rafael}. The finiteness of this entropy then
relies on UV regulators. Bianchi's beautiful idea \cite{bianchi} is to eliminate reliance on UV physics 
by considering {\em variations} of entanglement entropy.

Specifically, 4 dimensional classical gravity  
with matter is treated as a perturbation expansion in (the square root of) Newton's constant $G$ and 
the theory is quantized order by order in this expansion. This quantization is then applied to
the context of the `asymptotic  Rindler Horizons' defined by Jacobson and Parentani \cite{tp}.
If all fields are in the vacuum state,  the metric is just the flat metric with inertial coordinates $(t,x,y,z)$
and a Rindler wedge can be defined
as the region bounded by the union of the two half  hyperplanes $v=t+z=0,\; z\geq 0$ and $u=t-z=0, \;z\geq 0$. 
The fields are then varied about the vacuum 
so as to describe a stress energy  pulse which hits the $u=0$  boundary of the  wedge at some $t>0$. 
The Einstein equations imply that this stress energy perturbs the flat geometry\footnote{\label{nograviton}We shall restrict attention to a perturbation only of the matter vacuum so that the graviton 
is still in its vacuum.}.  
A perturbed Rindler horizon can then be defined  as a set of light rays which are initially divergent  at $t=0=v$ 
and are then  focussed
by the matter so that they emerge collimated at future null infinity.

The first order variation of the entanglement entropy is given by equation (\ref{s1st})
where the spatial region of interest, $B$, is now the half-space $t=0, z\geq 0$.    
The UV contribution of the first term is computed across the surface at $t=z=0$. 
The (expectation value) 
geometry of the unperturbed flat metric and the perturbed one agree to leading order in $\G$ in the vicinity of this 
surface (see equation (\ref{ht}) below). Since the UV contribution is only sensitive to the spacetime geometry 
in the vicinity of this surface, one
expects no `UV' component to this variation. This is the key idea of Bianchi which allows the possibility 
of computing the entropy variation without recourse to UV physics.

The second term in equation (\ref{s1st}) is then related to (as before, $1/\hbar$ times an integral of) 
the second term of equation (\ref{t1st}) 
through the First Law of Entanglement Entropy.
\footnote{Reference \cite{bianchi} predates  Reference \cite{myersetal} so that Bianchi derived the relation between
entanglement entropy  variation and stress energy  independent of Reference \cite{myersetal}.}
The  quantum  Einstein equations at leading order in $\sqrt{G}$ (see (\ref{ht}) below) 
then  relate this second term of equation (\ref{t1st}) with 
the first order variation of the area of the horizon. This area 
variation corresponds to  the first order variation of the Bekenstein Hawking entropy.

What about the possible IR contributions from the first term, of equation (\ref{s1st})?
\footnote{As in Reference \cite{ted}, there is no explicit mention of possible IR contributions to the first term in equation (\ref{s1st}) in 
Reference \cite{bianchi}; for our derivation of Bianchi's results to go through,
 these contributions should be neglegible. As shown below, this  is true if {\bf A2} holds.\label{fnbianchi}}
While an appeal to assumption {\bf A1} of section 4 allows for their neglect, we make the case
for their neglect stronger by arguing below that any contributions arising from derivatives 
with respect to `$\lambda$'(i.e. from metric change) are higher order in the perturbative 
parameter $G$ and can hence be dropped. Our argument applies to any term with `$\lambda$' derivatives
in equations (\ref{s1st}), (\ref{s1}) and (\ref{s2}).

The first step in the argument is to emphasise that the contribution to the entropy variation arising 
from the second term of equation (\ref{s1st}) is {\em independent} of the perturbative expansion parameter $G$.
This contribution evaluates to the area variation {\em divided} by $G$; equation (\ref{ht}) shows that the
area variation itself is of order $G$ so that this variation divided by $G$ is indeed independent of 
$G$ (and equal to an appropriate  $G$-independent integral of the stress energy via the First Law of 
Entanglement Entropy). The second step is to note again that the beauty of Bianchi's \cite{bianchi} set up
is that there is  no geometry change in the vicinity of the entangling surface so that one expects all
contributions to the entropy variation to be UV finite. Specifically there is no dependence on a UV cutoff
at the Planck Scale and hence no dependence on $G^{-1}$. Further, if we take $G\rightarrow 0$,  equation (\ref{ht})
implies that there is no geometry change. Together with the UV finiteness property, this is in turn implies that any derivative of the entropy with respect to $\lambda$
must vanish in the $G\rightarrow 0$ limit.

\newpage

To summarise: In addition to the arguments presented in support of {\bf A1}, the argument above implies the
 statement {\bf A2}:\\
{\em All terms with $\lambda$ derivatives in equations (\ref{s1st}), (\ref{s1}) and (\ref{s2}) yield contributions
which vanish as $G\rightarrow 0$ and can be dropped in comparision with the leading order $G$-independent
contribution to the entropy which is of concern to us.}

In particular, this implies the neglect of IR contributions to the first term of (\ref{s1st}) so that 
Bianchi's results on the equality of the first order variation of entanglement entropy with that of the 
Bekenstein- Hawking area variation follow.

One may question as to what happens when the matter is in a coherent state which is a finite, small variation about the 
vacuum; on the one hand it seems that  one can throw
stress energy through the horizon with no entropy cost but on the other, the Einstein equations mandate that
any energy flux must cause area change through the Raychaudhuri equations. 
The entanglement entropy change, as in section 5, vanishes to first order in the smallness parameter
under our  assumption {\bf A2} above. As shown in section 2 the stress energy contribution
to the right hand side of equation (\ref{ht}) also vanishes to 
to first order in the smallness parameter 
so that 
there is no area change to first order in the smallness parameter consistent with no change in the entanglement entropy.
Thus with the caveats that our assumption {\bf A2} on IR contributions to equation (\ref{s1st}) holds  
we see that there is no contradiction with the first order analysis of Reference \cite{bianchi}.

What about an analysis to second order in the smallness parameter for coherent states? 
In the setting of References \cite{bianchi,bianchisatz}, since we are using the 
quantum Einstein equations 
(to leading order in $\sqrt{G}$), rather than attempting to derive them as in Reference 
\cite{ted}, we can proceed
further as follows.
%

At the classical level, the key equations  \cite{bianchisatz} which relate the `Coulombic'
part of the gravitational perturbation at leading order in $G$ to the stress energy are:
\ba
g_{\mu \nu} &= &\eta_{\mu \nu} + \G(h^{{\rm rad}}_{\mu \nu} + h^{{\rm source}}_{\mu \nu})\\
h^{{\rm source}}_{\mu \nu}&=& -\G\int d^4x^{\prime}G_R(x, x^{\prime})
(T_{\mu \nu}(x^{\prime})-  \frac{1}{2}\eta_{\mu \nu}T^{\rho}_{\rho}(x^{\prime}))
\label{ht}
\ea
where $h^{{\rm rad}}_{\mu \nu}$ is a solution to the massless wave equation in the flat spacetime, $G_R(x,x^{\prime})$
is the retarded Green's function for this equation,  
and $T_{\mu \nu}$ is the total stress energy of the matter field $\Phi$ and the gravity wave
$h^{{\rm rad}}_{\mu \nu}$ .
In the quantum theory,  this stress energy is replaced by the corresponding normal ordered (with respect to the 
flat spacetime) operator. Since we are interested in the case where the graviton state remains in its vacuum 
(see Footnote \ref{nograviton}),
when we perturb the matter state away from its vacuum 
the expectation value metric (which we also denote by $g$) only gets contributions from that of 
the ${\hat h}^{{\rm source}}_{\mu \nu}$ operator.

Let us now focus in detail 
on perturbations to coherent states of the matter. Accordingly we define the  1 parameter set of 
coherent states $\psi (\lambda ):= |\lambda H \ket$ (see equation ({\ref{mdef})).
Taking the expectation value of  equation (\ref{ht}) in the state $\psi (\lambda )$  defines the 1 parameter
family of corresponding expectation value metrics $g(\lambda )$. 
It follows immediately that  the leading order geometry change $g(\delta ) - \eta$ is  second order in $\delta$,  
first order in $G$ and zeroth order in $\hbar$ (the latter follows from equation (\ref{mhad})).

As discussed above, the variation of 
 the entanglement entropy vanishes 
(modulo  assumption {\bf A2}) to first order in the smallness parameter $\delta$ as does the 
 stress energy contribution to the right hand side of (\ref{ht}).
We now examine the second order contributions (\ref{s2}).
The first term of (\ref{s2}) corresponds to the contribution from 
the varied geometry and the unvaried (vacuum) state. Its IR part can be neglected by our assumption {\bf A2}.
The  UV part of the first term vanishes, similar to the vanishing of the 
UV part of the first term of (\ref{s1}) discussed above, 
because the geometry in the vicinity of the entangling surface does not 
change to leading order in ${G}$. 
The second term in (\ref{s2}) has no $\lambda$ derivative. Therefore, it corresponds to 
the second order contribution from the varied (coherent) state on the unvaried (flat) geometry. This
contribution   vanishes identically by virtue of the equality (see section 2) 
of the vacuum and coherent state entanglement entropies 
on the unvaried geometry. The last `mixed derivative' term in (\ref{s2}) has a single derivative with respect to
$\lambda$ and can be dropped by Assumption {\bf A2}. In addition the following formal evaluation of this
term using the chain rule suggest that it is not only higher order in $G$ but also vanishes identically:
\be
\big( \frac{\partial^2 S}{\partial \lambda \partial \tau} \big)_{\lambda=\tau=0}
=  
\left( \int_{}\frac{dg(x)}{d\lambda} \left(\frac{\delta}{\delta g_(x)}
\left(\frac{\partial S}{\partial \tau} \right)\right)\right)_{\lambda=\tau=0} =0
\label{dg}
\ee
where we have used the fact (see discussion above) that the metric perturbation is of order $\delta^2$ so that 
$\frac{dg(x)}{d\lambda}|_{\lambda =0}=0$. It is not clear to us as to what the exact range of integration should be 
in equation (\ref{dg}) as this depends on the details of the way in which the state $\psi (\tau )$ is identified
as a state on the perturbed spacetime geometry. Nevertheless it seems reasonable that, given that the 
metric is unchanged to first order, any single derivative with respect to $\lambda$ vanishes when evaluated at the 
origin. In any case both assumption {\bf A2} and the formal evaluation (\ref{dg}) indicate that the mixed derivative
term can be dropped.


The estimate above, finds a vanishing  entanglement entropy variation when the entangling surface is the noncompact
region $(z=t=0)$
and implies that 
the entanglement entropy per unit area of the entangling surface does not change to 
order $\delta^2$.

In contrast, as in \cite{bianchi}, equation (\ref{ht}) can be used to relate the change 
in ${G}^{-1}$ times the area of a 
small part   
of the $t=0,z=0$ surface (representing the unperturbed cross section of the null congruence of interest)
to the stress energy flux so that $G^{-1}$ times the area change is {\em nontrivial} to order $\delta^2$.
We emphasise that, as seen from equation (\ref{ht}), the area change is itself of order $G$ so that this change
divided by $G$ is independent of $G$ and proportional to $\delta^2$ (which is exactly the order of the coherent state
stress energy on the right hand side of equation (\ref{ht})). Thus on the one hand the $G$ -independent 
entanglement entropy change
for coherent states vanishes to order $\delta^2$ and on the other, the $G$-indepenent Bekenstein Hawking entropy change 
is non-trivial to order $\delta^2$.

Hence if the formal aspects of the above arguementation (including the issue of the existence of a Taylor
series expansion as envisaged in equations (\ref{s1s2})- (\ref{s2})) can be made rigorous and if assumption 
{\bf A2} holds 
\footnote{See Footnote \ref{fnbianchi} in this regard.}
%
then  the ideas of \cite{bianchi} do not seem to apply to the 
coherent state example of section 2.
%

One viewpoint 
\footnote{We thank the referee for bringing this viewpoint to our notice.}
on this result is that  coherent state excitations do not represent 
quasitatic, near equilibrium perturbations for which the Clausius relation ``$dQ/T= dS$''
is expected to hold. If this viewpoint is correct then one may argue that the near equilibrium
changes which underly  black hole thermodynamics  are simply not represented by 
 coherent state perturbations. Indeed, the First Law of Entanglement Entropy 
 relates the change in entanglement entropy `$dS$'  to the boost energy influx (represented by the 
 stress energy)  divided by a `geometric' temperature proportional to $\hbar$ and may be
 viewed as a quantum version of the Clausius relation. From this point of view,  coherent states
 do not conform to this Clausius relation and hence one does not expect
 black hole thermodynamics to follow.
 
 We do not subscribe to  this viewpoint  for the following reasons.
First, it is not clear to us that the change in black hole entropy can be accounted for by entanglement entropy 
change alone. If we are open to the possibility that the two are not identical, the `$dS$' term in the 
First Law of Entanglement Entropy need not coincide with the change in thermodynamic entropy so that the latter
could still be driven by the physical stress energy even if this stress energy is non-trivial only at second order.
In other words we are
unsure if the First Law of Entanglement Entropy is indeed a quantum version of the Clausius relation for 
black hole thermodynamics. 
Second,  a coherent state excitation is as close to  a classical excitation as possible.
 The coherent state field expectation value as well as its stress energy can 
be nicely confined to a region of compact support  on the $t=0$ slice  so as not to extend across the horizon at $t=0$.
 More in detail the coherent state can 
  be modelled on a classical function which is dominated by
 large wavelengths and whose support is confined away from the horizon at $t=0$ and thereby, 
on intuitive physical grounds, constitutes a small, slowly evolving
 quasistatic perturbation moving towards the horizon.  Further support for the nice properties of coherent states 
 comes from the work of Ford and Kuo \cite{fordkuo}, in which it is argued that stress energy 
 fluctuations for coherent states are minimal.  
On physical grounds, one would  expect near equilibrium perturbations to be modeled by states which do 
exhibit such minimal  fluctuations of stress energy and, consequently, of geometry.
  Indeed for a wide variety of states which constitute nontrivial first order perturbations
 of the vacuum and  the stress energy, Ford and Kuo \cite{fordkuo} argue that  there are regions in 
spacetime where the stress energy fluctuations are large.
 Such states include single mode superpositions of the vacuum and a 2 particle state as well as single mode squeezed
 states which are far away from coherent states in the sense that the squeezing parameter 
 is larger than the coherent state parameter \cite{fordkuo}.  
 \footnote{Single mode states extend all over spacetime and hence are 
 not valid examples of  small, confined  perturbations. However,  it seems reasonable to us to  expect 
 that  similar results hold for confined, wave packet generalizations of such states which are dominated
 by some large wavelength; unfortunately, we do not know of explicit examples and it would be of interest to
investigate if such examples exist.}
 In general the expectation is that a state exhibits large fluctuations of stress energy wherever
 the stress energy is negative \cite{fordkuo}.  Coherent states have everywhere positive
 energy density (for classical field theories which satisfy positive energy conditions as assumed
 in this work) and have minimal stress energy fluctuations everywhere. Hence in all these aspects
 we view them as ideal models for quasistatic near equilibrium perturbations.

 To summarise: We envisage at least two different viewpoints on our result that 
 the considerations of Reference \cite{bianchi}   for first order variations of 
 entanglement entropy do not seem to extend to the case of finite but small variations to coherent
 states. From the first point of view coherent state perturbations are simply not near equilibrium
 perturbations and it is no surprise that the considerations of Reference \cite{bianchi}  do
 not extend to this case as this case does not represent a physically relevant near equilibrium
 situation. The second viewpoint asserts that  coherent state perturbations 
 modeled on suitable classical solutions
 can indeed  be thought of as quasistatic, near equilibrium perturbations for which 
 fluctuations in the stress energy and metric are minimal. From this point of view, 
 our results indicate that such coherent state perturbations constitute a genuine obstruction to an 
 extension of the ideas of Reference \cite{bianchi} from the case of first order variations to the case of small but finite variations.

\section*{Acknowledgements:}
I thank Ted Jacobson for his generous 
help with my numerous questions with regard to Reference \cite{ted} and for his comments on a
draft version of this work. 
I thank  Abhay Ashtekar for discussions and an anonymous referee for her/his comments.

\appendix
\section{Coherent states in curved spacetime}
Let $(M,g)$ be a globally hyperbolic $d$- dimensional spacetime with metric $g$. 
Denote a Cauchy slice in $(M,g)$
by $\Sigma$ where $\Sigma$ is a $d-1$ dimensional orientable manifold without boundary. 
Let $\R_I,\R_{II}$ be closed 
 $d-1$ dimensional submanifolds of $\Sigma$ such that 
$\Sigma = \R_{I} \cup \R_{II}$  and $\R_{I}\cap \R_{II}=\S$ where $\S= \partial \R_{I} =\partial \R_{II}$ 
is a $d-2$ dimensional hypersurface.
We restrict attention to the case  that $\R_I$ is a small geodesic ball \cite{ted}
and the case $M=R^4$ with $\R_I$ the right half plane $z\geq 0$ \cite{bianchi}. 

Let $\Phi$ be a free  scalar field propagating on $(M,g)$. We assume that the free field dynamics on $M,g$ 
exhibits unique evolution from initial data in any causal domain  and that smooth data of compact
support evolve to smooth solutions as in Lemma 3.1  of \cite{wald}.
Let the initial data on $\Sigma$ be  $(\phi$, $\pi)$ where $\phi$  denotes the scalar field on $\Sigma$ and $\pi$
its conjugate momentum. Let $\Phi$ be quantized with respect to some choice of 
complex structure $J$ (i.e. mode decomposition) \cite{wald} of this data on $\Sigma$ and let 
$|0\ket$ be the vacuum state and $\F$ the Fock space, with respect to this
choice. 
Likewise, considering $\R_{I}$ and $\R_{II}$ as Cauchy slices for their domains of dependence $D_I, D_{II}$, let 
$J_I, J_{II}$ be complex structures for data on $\R_I, \R_{II}$ with associated Fock
quantizations $\F_I, \F_{II}$. 

Since the field operators  ${\hat \phi}(x), {\hat \pi}(x)$ are operator valued distributions on $\Sigma$,
given smooth smearing functions $e, f$  of compact support  on $\Sigma$ we have that the operator
${\hat U}(e,f)$ is unitary where 
\be
 {\hat U}(e,f):= e^{i\int_{\Sigma} f(x){\hat \phi} (x) - e(x) {\hat \pi} (x)}  
\ee
Define the coherent state $|e,f\ket$ as
\be
|e,f\ket := {\hat U}(e,f) |0\ket
\ee

Define ${\hat \rho}_I(e,f) = \tr_{\F_{II}}(|e,f\ket \bra e,f|)$  
so that ${\hat\rho}_I (0,0)= \tr_{\F_{II}}(|0\ket \bra 0|)$. We argue  below that the entanglement entropy
$S(e,f)= -\tr_{\F_I}({\hat \rho}_I(e,f)\ln {\hat \rho}_I(e,f))$ is the same as the vacuum entanglement 
entropy $S(0,0)= -\tr_{\F_I}({\hat \rho}_I(0,0)\ln {\hat \rho}_I(0,0))$ modulo issues of UV divergences.

Define 
\ba
{\hat U}_I(e_I,f_I) & =& e^{i\int_{\R_I} f_{I} (x)\phi (x) + e_{I}(x) \pi (x)},\\
e_I= e \;\;\;{\rm on} \;\;\;\R_I & &e_I=0 \;\;\;{\rm on}\;\;\;\R_{II}.
\\
f_I= f \;\;\;{\rm on} \;\;\;\R_I & &f_I=0 \;\;\;{\rm on}\;\;\;\R_{II}.
\ea
and likewise for $I\leftrightarrow II$. 

Let $e,f$ be supported away from $\S$. Then we have that ${\hat U}_I(e_I,f_I)$ and ${\hat U}_{II}(e_{II},f_{II})$  
are unitary operators on $\F, F_I$ and $\F, F_{II}$ respectively and that 
\be
{\hat U}(e,f)= 
{\hat U}_{II}(e_{II},f_{II}){\hat U}_{I}(e_I,f_I)= {\hat U}_{I}(e_{I},f_{I}){\hat U}_{II}(e_{II},f_{II})
\label{u=u1u2}
\ee
We have that 
\ba
{\hat \rho}_I(e,f) &=& \tr_{\F_{II}} ({\hat U}_{II}(e_{II},f_{II}){\hat U}_{I}(e_I,f_I)|0\ket\bra 0|
             {\hat U}_{I}(e_I,f_I)^{\dagger}{\hat U}_{II}(e_{II},f_{II})^{\dagger})\\
&=& \tr_{\F_{II}}({\hat U}_{I}(e_I,f_I)|0\ket \bra 0|
             {\hat U}_{I}(e_I,f_I)^{\dagger}) = 
{\hat U}_{I}(e_I,f_I)\tr_{\F_{II}}(|0\ket \bra 0|){\hat U}_{I}(e_I,f_I)^{\dagger}
\ea
where we have used the invariance of the Trace under unitary tranformations, and the fact that $U_{I}(e,f)$ acts
as the identity operator on  $\F_{II}$ in the last  line. The equality of $S(e,f)$ and $S(0,0)$ then follows, once again
from the invariance of the Trace under unitary transformations.
Our demonstration suffers from exactly the same UV divergences as in the case of flat spacetime discussed in section 2.   
That discussion clearly applies here and, while we do not provide an explicit regularization, indicates the 
any UV regulation softens the sharp boundary ${\cal S}$ and strongly suggests that the above argument can be made
well defined.

Next, let the vacuum state be Hadamard. 
It is straightforward to check, for example using sections 3,4 of \cite{wald}, that 
the two point function in the coherent state $|e,f\ket$ evaluates to
\be 
\bra e,f| {\hat \Phi}(X){\hat \Phi}(X^{\prime})|e,f\ket=
\bra 0| {\hat \Phi}(X){\hat \Phi}(X^{\prime})|0\ket + H(X)H(X^{\prime})
\label{had}
\ee
where $H$ is scalar field solution obtained from the initial data   $(e=\phi (x), f=\pi (x)), \; x\in \Sigma$ 
and $H(X), H(X^{\prime})$ are the values of $H$ at the spacetime points $X, X^{\prime} \in M$.
From our assumptions on free field evolution it follows that $H$ is smooth. Equation (\ref{had}) then implies 
that $|e,f\ket$ is also Hadamard for any choice of smooth and compactly supported $(e,f)$.
Since we are interested in stress energy expectation value
relative to the vacuum and since the stress energy operator is quadratic in the field, 
we adopt a normal ordering prescription with respect to our choice of annihilation and creation operators on $\F$.
We are interested in coherent states which are first order departures from the vacuum. For 
some small positive parameter $\delta$, consider the coherent state $|\delta e, \delta f\ket$ so that
\be
 |\delta e, \delta f\ket = e^{i\delta \int_{\Sigma} f(x){\hat \phi} (x) - e(x) {\hat \pi} (x)}|0\ket
\label{def}
\ee
Expanding the exponential out and noting that the field operators are linear in the annihilation- creation modes,
we obtain an expansion of the coherent state in terms of $n$- particle states with the norm of the $m$-particle
component of order $\delta^m$. In particular the one particle component is of order $\delta$ so that this 
coherent state constitutes a first order variation from the vacuum.
Since the (normal ordered) stress energy tensor is quadratic in the field operators, the order $\delta$ contribution
to its expectation value which is from the off diagonal 
matrix element between its vacuum component and its  1 particle 
component, vanishes,
so that the  leading order contribution is at {\em second} order 
\footnote{This can also be seen directly from the 
expression for the two point function (\ref{had}) which implies that the stress energy expectation value is exactly  
that of the classical solution $\delta \times H$.} 

More generally, since the stress energy operator is quadratic in the modes, the only way for any 
state to be both a first order variation of the vacuum and exhibit a non-trivial normal ordered 
stress energy expectation value
at first order is if its two particle component is of order $\delta$.

\end{document}